------------------------------------------------------------------

\documentstyle[aps,12pt]{revtex}
\begin{document}

\title{Quantum Three-Body Problem}

\author{Zhong-Qi Ma \thanks{Electronic
address:MAZQ@BEPC3.IHEP.AC.CN}}

\address{China Center for Advanced Science and Technology
(World Laboratory), Beijing 100080\\
and Institute of High Energy Physics, 
Beijing 100039, The People's Republic of China}

\author{An-Ying Dai}

\address{Department of Applied Physics, Beijing Institute of Technology, 
The People's Republic of China}


\maketitle

\vspace{5mm}
\begin{abstract}
An improved hyperspherical harmonic method for the quantum 
three-body problem is presented
to separate three rotational degrees of freedom completely
from the internal ones. In this method, the Schr\"{o}dinger 
equation of three-body problem is reduced to a system of 
linear algebraic equations in terms of the orthogonal bases 
of functions. As an important example in quantum mechanics, 
the energies and the eigenfunctions of some states of the 
helium atom and the helium-like ions are calculated. 

\vspace{3mm}
\noindent
PACS number(s): 03.65.Ge, 31.15.-p, and 02.30.Gp

\end{abstract}

\maketitle

\section{INTRODUCTION}
Exact solutions played very important roles in the development of
physics. The exact solutions of the Schr\"{o}dinger equation for
a hydrogen atom and for a harmonic oscillator laid a foundation 
of quantum mechanics. The next simplest atom is the helium atom, 
which is a typical three-body problem and has not been well-solved 
both in classical mechanics [1] and in quantum mechanics [2]. 
The Faddeev equations [3] are popularly used in both scattering 
processes [4] and bound state calculations [5] for the quantum 
three-body problem. However, only a few analytically solvable 
examples were found [6]. For the helium atom, variational methods 
can achieve 9- to 12-place precision for energy values with a few 
hundred or thousand variational parameters [7-10]. There are many 
reasons to prefer the direct solutions of the three-body Schr\"{o}dinger
equation over the variational one, for example, the analytic structure 
of the variational wavefunction is chosen arbitrarily. The accurate
direct solution of the three-body Schr\"{o}dinger equation with
the separated center-of-mass motion has been sought based on
different numerical methods, such as the finite difference [11],
finite element [12], complex coordinate rotation [13], 
hyperspherical coordinate [14], and hyperspherical harmonic [15-17]
methods. 

In the hyperspherical harmonic method [15-17], the six Jacobi 
coordinates, after separation of the center-of-mass motion, 
are separated to one hyperradial variable $\rho$ and five 
hyperangular variables $\Omega$. The wave function is presented 
as a sum of products of hyperradial function and the hyperspherical 
harmonic function, depending on $\Omega$. Since three degrees 
of freedom for the rotation of the system are not separated 
completely from the internal ones, there is the huge degeneracy 
of the hyperspherical basis. The interactions in the three-body 
problem is not hyperspherically
symmetric so that the matrix elements of the potential have to be 
calculated between different hyperspherical harmonic states. 
Another difficulty in the practical calculations by the 
hyperspherical harmonic method is the slow convergence of the 
series. The convergence was fastened by decomposing the wavefunction 
$\psi=\chi \phi$, where $\chi$ is chosen to take into account 
the singularities of the potential and the clustering properties 
of the wavefunction, and $\phi$ is the part to be expanded in 
hypersphericals [15]. 

In an unpublished paper [18], Hsiang and Hsiang devised a 
principle method to reduce the Schr\"{o}dinger equation of 
three-body problem to a system of linear algebraic 
equations. Although their method was incomplete, it contained 
a good idea for simplifying the quantum three-body problem. Developing 
their idea, we suggest an improved hyperspherical harmonic method to 
separate three rotational degrees of freedom completely from the 
internal ones, and the decomposition of the function for fastening the 
convergence of the series in the hyperspherical harmonic method
[15-17] is still effective in the improved method.

First, after the separation of the center-of-mass motion and 
the whole rotation, the three-body Schr\"{o}dinger equation 
with any given angular momentum and parity is reduced to the 
coupled partial differential equations, depending upon only 
three variables, which are invariant in the rotation of the system. 
The coupled equations are closed because the potential depends 
only upon those three variables due to its symmetry in the SO(3) 
rotation and the space inversion.
Then, we expand the solutions of the coupled equations by the 
orthogonal bases of functions to reduce the coupled partial 
differential equations to the coupled ordinary differential 
equations. The bases of functions are similar to those in the 
$S$ wave case of the hyperspherical harmonic method [16], but 
in our method they are effective for any given angular momentum. 
The coupled ordinary differential equations are similar to the 
hyperradial equations for the hyperradius in the hyperspherical 
harmonic method [16], but there is an important difference between 
them. The differential operators in each ordinary differential 
equation of our coupled equations are same so that the equations 
can be reduced to a system of linear algebraic equations by the 
spectral decomposition of the hyperradial function. 
This improved hyperspherical harmonic method is effective 
for the three-body problems of both identical and 
non-identical particles, and effective for any pair potential, 
which depends only upon the distance of each pair of particles. 

The improved hyperspherical harmonic method greatly decreases 
the calculating time, because to solve a system of linear algebraic 
equations even with the equation number more than 1000 is faster 
than to solve a system of ordinary differential equations with the 
equation number less than ten. Furthermore, all the technique used 
in the hyperspherical harmonic method to fasten the convergence
of the series can also be used in the improved hyperspherical 
harmonic method.

In this paper we will demonstrate the improved hyperspherical 
harmonic method in some detail. In Sec. II we introduce 
our notations and write the Schr\"{o}dinger equation for 
a three-body problem in the center-of-mass frame by the Jacobi
coordinates. Due to the $SO(3)$ symmetry of the Schr\"{o}dinger 
equation, in Sec. III we will explain how to separate the 
rotational degrees of freedom completely from the internal ones, and
derive the formulas for reducing the three-body Schr\"{o}dinger 
equation with any given angular momentum
and parity to a system of partial differential equations.
In Sec. IV we introduce three new variables through the complex 
vector coordinates [19]. By a complete set of normalized and 
orthogonal bases of functions of those new variables, the 
Schr\"{o}dinger equation for any angular momentum and parity 
is further reduced to a system of linear algebraic 
equations with some coefficients to be calculated. In terms of some 
identities, proved in Appendix, the alternating series contained in the 
formulas for the coefficients are calculated analytically.
As an important example in quantum mechanics, the problem 
of a helium atom is studied in some detail. We solve the
system of linear algebraic equations for the
lowest-energy states of the $S$ wave and the $P$ wave 
in both the parahelium and the orthohelium in Sec. V, 
respectively. We also calculate the energies of the ground
state of some helium-like ions there. The solution of the 
Schr\"{o}dinger equation for the three-body problem is called exact
in the meaning that it is expressed as an infinite series in terms of 
the orthogonal bases of functions, and can be calculated in any 
precision if enough terms in the series are included. Due to
our limited condition in computer, in the present paper we 
still use the truncation method of the series and do not fasten 
the convergence of the series by decomposition of the wavefunction,
so that the calculating precision is in the thousandths for the
energy of the ground state of the helium atom and in the hundredths 
for the remaining calculating results. The calculating precision 
will be improved elsewhere. The main purpose of this paper is to 
present the improved hyperspherical harmonic method. Some discussions are 
given in Sec. VI.

\section{THE SCHR\"{O}DINGER EQUATION}

Denote by ${\bf r_{j}}$ and by $M_{j}$, $j=1,2,3$, the position 
vectors and the masses of three particles in a three-body problem,
respectively. The relative masses are $m_{j}=M_{j}/M$, where 
$M$ is the total mass, $M=\sum M_{j}$. The Schr\"{o}dinger 
equation for the three-body problem is
$$- (\hbar^{2}/2M) \bigtriangleup \Psi +V \Psi =E \Psi , \eqno (1) $$
$$\bigtriangleup = \displaystyle \sum_{j=1}^{3}~\displaystyle
m_{j}^{-1} \bigtriangleup_{r_{j}}  , \eqno (2) $$

\noindent
where $\bigtriangleup_{r_{j}}$ is the Laplace operator with respect 
to the position vector ${\bf r}_{j}$
$$\bigtriangleup_{r_{j}}=\displaystyle {\partial^{2} \over 
\partial r_{j1}^{2} }+ \displaystyle {\partial^{2} \over 
\partial r_{j2}^{2} } +\displaystyle {\partial^{2} \over 
\partial r_{j3}^{2} }, \eqno (3) $$

\noindent
and $V$ is a pair potential, depending only upon the distance
of each pair of particles. For definiteness, we discuss the Coulomb
potential:
$$V=\displaystyle {Z_{2}Z_{3}e^{2}\over |{\bf r_{2}}-{\bf r_{3}}|}+
\displaystyle {Z_{3}Z_{1}e^{2}\over |{\bf r_{3}}-{\bf r_{1}}|}+
\displaystyle {Z_{1}Z_{2}e^{2}\over |{\bf r_{1}}-{\bf r_{2}}|}, 
\eqno (4) $$

\noindent
where $Z_{j} e$ denotes the electric charge of the $j$th particle.

Now, we replace the position vectors ${\bf r}_{j}$ by the Jacobi
coordinates ${\bf R}_{j}$:
$${\bf R}_{1}=m_{1}{\bf r}_{1}+m_{2}{\bf r}_{2}+m_{3}{\bf r}_{3},
~~~~~{\bf R}_{2}={\bf x}+\sqrt{\displaystyle {m_{1}\over m_{2}+m_{3}}}
~ {\bf R}_{1},~~~~~{\bf R}_{3}={\bf y},
\eqno (5) $$
$${\bf x}=-\sqrt{\displaystyle {m_{1}\over m_{2}+m_{3}}}~{\bf r_{1}},~~~~~
{\bf y}=\sqrt{\displaystyle {m_{2}m_{3}\over m_{2}+m_{3}}}
~\left({\bf r_{2}-r_{3}}\right), 
\eqno (6) $$

\noindent
where ${\bf R}_{1}$ describes the position of the center-of-mass and 
vanishes in the center-of-mass frame. ${\bf x}$ and ${\bf y}$ 
describe the position vectors of three particles in the 
center-of-mass frame for two special cases, respectively: 

i) The second particle coincides with the third particle, 
$${\bf r}_{1}=-\sqrt{\displaystyle {m_{2}+m_{3}\over m_{1}}}~{\bf x},
~~~~{\bf r}_{2}={\bf r}_{3}
=\sqrt{\displaystyle {m_{1}\over m_{2}+m_{3}}}~{\bf x},~~~~
\displaystyle \sum_{j=1}^{3}~m_{j}{\bf r}_{j}^{2}={\bf x}^{2}.
\eqno (7) $$

ii) The first particle is located at the origin (the center-of-mass), 
$${\bf r}_{1}=0,~~~~m_{2}{\bf r}_{2}=-m_{3}{\bf r}_{3}
=\sqrt{\displaystyle {m_{2}m_{3}\over m_{2}+m_{3}}}~{\bf y},~~~~
\displaystyle \sum_{j=1}^{3}~m_{j}{\bf r}_{j}^{2}={\bf y}^{2}.
 \eqno (8) $$

A straightforward calculation by replacement of variables shows 
that the Laplace operator in Eq. (2) and the angular momentum 
operator ${\bf L}$ are directly expressed with respect to ${\bf R}_{j}$:
$$\bigtriangleup = \displaystyle \sum_{j=1}^{3}~\bigtriangleup_{R_{j}}, $$
$${\bf L}=-i\hbar \displaystyle \sum_{j=1}^{3}~ {\bf r_{j}}
\times \bigtriangledown_{r_{j}}
=-i\hbar \displaystyle \sum_{j=1}^{3}~ {\bf R_{j}}
\times \bigtriangledown_{R_{j}}, $$

\noindent
In the center-of-mass frame, they further reduce to
those with respect to the position vectors ${\bf x}$ 
and ${\bf y}$:
$$\bigtriangleup = \bigtriangleup_{x}+ 
\bigtriangleup_{y},~~~~~
{\bf L}=-i\hbar {\bf x}\times \bigtriangledown_{x}
-i\hbar {\bf y}\times \bigtriangledown_{y}. \eqno (9) $$

\section{EIGENFUNCTIONS OF THE ANGULAR MOMENTUM}

The Schr\"{o}dinger equation (1) is spherically symmetric so 
that its solution can be factorized into a product of an
eigenfunction of the angular momentum ${\bf L}$ and a
"radial" function, which only depends upon three variables,
invariant in the rotation of the system:
$$\xi_{1}={\bf x\cdot x},~~~~~
\xi_{2}={\bf y\cdot y},~~~~~
\xi_{3}={\bf x\cdot y}. \eqno (10) $$

\noindent
We call them the "radial" variables in this paper.

For a particle moving in a central field, the eigenfunction of 
the angular momentum is the spherical harmonic function 
$Y^{\ell}_{m}(\theta, \varphi)$. What is the generalization
of the spherical harmonic function to the three-body problem?

A natural idea for generalization is to introduce the angular 
variables, for example, the Euler angles. In this way, as 
discussed by Wigner (see p.214 in [20]), the eigenfunction of 
the angular momentum is the representation matrix of the $SO(3)$ 
group, $D^{\ell}_{mm'}(\alpha,\beta,\gamma)$. However, the 
Schr\"{o}dinger equation with respect to the Euler angles is
singular. In the hyperspherical harmonic method [16] the 
spherical harmonic function is generalized to the hyperspherical
harmonic function, which depends upon five angular variables 
$\Omega=(\alpha, \theta_{x}, \varphi_{x}, \theta_{y}, \varphi_{y})$, 
where $\theta_{x}$ ($\theta_{y}$) and $\varphi_{x}$ ($\varphi_{y}$) 
are the angular variables of ${\bf x}$ (${\bf y}$). Obviously, the 
rotational degrees of freedom are not separated from the internal 
ones. Is it necessary to introduce the angular variables explicitly?

In order to avoid the angular variables, let us study the
other properties of the spherical harmonic function, in addition 
to the eigenfunction of the angular momentum. As is well known, 
${\cal Y}^{\ell}_{m}({\bf x})\equiv r^{\ell}Y^{\ell}_{m}(\theta, \varphi)$,
where $(r,\theta,\varphi)$ are the spherical coordinates for
the position vector ${\bf x}$, is a homogeneous polynomial of 
degree $\ell$ with respect to the components of ${\bf x}$, 
and satisfies the Laplace equation as well as the eigen-equation 
for the angular momentum, but does not contain the angular 
variables and the $r^{2}$ factor explicitly. The number of 
linearly independent homogeneous polynomials of degree $\ell$ 
with respect to the components of ${\bf x}$ is $N(\ell)$:
$$N(\ell)=\displaystyle \sum_{s=0}^{\ell}(\ell-s+1)=
(\ell+1)(\ell+2)/2. $$  

\noindent
Removing those homogeneous polynomials containing the factor
$r^{2}={\bf x\cdot x}$, we obtain that the number of linearly 
independent homogeneous polynomials of degree $\ell$, not containing 
the factor $r^{2}$, is 
$$N(\ell)-N(\ell-2)=2\ell+1. \eqno (11) $$

\noindent
It is nothing but the number of ${\cal Y}^{\ell}_{m}({\bf x})$
with the same angular momentum $\ell$.

Now, for the three-body problem, there are three radial variables
$\xi_{j}$ in the center-of-mass frame. 
${\cal Y}^{q}_{m}({\bf x}){\cal Y}^{\ell-q}_{m'}({\bf y})$
are the linearly independent homogeneous polynomials of degree $\ell$
with respect to the components of the position vectors ${\bf x}$ 
and ${\bf y}$, not containing the factors $\xi_{1}$ and 
$\xi_{2}$. The number of those polynomials is $M(\ell)$: 
$$M(\ell)=\displaystyle \sum_{q=0}^{\ell}~(2q+1)(2\ell-2q+1)=
(\ell+1)(2\ell^{2}+4\ell+3)/3. $$  

\noindent
Removing those polynomials containing the factor $\xi_{3}$, 
we obtain that the number of linearly independent 
homogeneous polynomials of degree $\ell$, not containing the 
factors $\xi_{j}$, is 
$$M(\ell)-M(\ell-2)=4\ell^{2}+2. \eqno (12) $$

According to the theory of angular momentum [20], the polynomials 
${\cal Y}^{q}_{m}({\bf x}){\cal Y}^{\ell-q}_{m'}({\bf y})$
can be combined to be the eigenfunctions of ${\bf L}^{2}$ by 
the Clebsch-Gordan coefficients. The number of the combinations 
with the angular momentum $\ell$ and $(\ell-1)$, which do not 
containing the factors $\xi_{3}$, is 
$$(2\ell+1)(\ell+1)+(2\ell-1)(\ell-1)=4\ell^{2}+2.  $$

\noindent
It coincides with the number given in Eq. (12). In other words, 
the eigenfunctions of the angular momentum ${\bf L}^{2}$ with
the eigenvalue $\ell(\ell+1)$, not containing the factors $\xi_{j}$,
are those homogeneous polynomials of degree $\ell$ or degree $(\ell+1)$.
Let us introduce a parameter $\lambda=0$ or $1$ to identify them.
Due to the property of the spherical symmetry, what we need
is to write the eigenfunctions of ${\bf L}^{2}$ with the largest 
eigenvalue of $L_{3}$, where the normalization factors do not 
matter with us. Denote them by $Q_{q}^{\ell \lambda}({\bf x,y})$ 
with the degree $(\ell+\lambda)$, $\lambda \leq q \leq \ell$. 
$$\lambda=0:~~~~~
{\cal Y}^{q}_{q}({\bf x}){\cal Y}^{\ell-q}_{\ell-q}({\bf y}),
~~~~~~~~~~~~~~~~~~~~~~~~~~~~~~~~~~~~$$
$$\lambda=1:~~~~~\sqrt{\ell-q+1}{\cal Y}^{q}_{q-1}({\bf x})
{\cal Y}^{\ell-q+1}_{\ell-q+1}({\bf y})-\sqrt{q}{\cal Y}^{q}_{q}({\bf x})
{\cal Y}^{\ell-q+1}_{\ell-q}({\bf y}).
 \eqno (13) $$

\noindent
Since ${\cal Y}^{\ell}_{\ell}({\bf x})\sim \left(x_{1}+ix_{2}\right)^{\ell}$ 
and ${\cal Y}^{\ell}_{\ell-1}({\bf x})\sim -\sqrt{2\ell} 
\left(x_{1}+ix_{2}\right)^{\ell-1}x_{3}$, we have
$$Q_{q}^{\ell \lambda}({\bf x,y})=\{(q-\lambda)!(\ell-q)!\}^{-1}
(x_{1}+ix_{2})^{q-\lambda}(y_{1}+iy_{2})^{\ell-q} $$
$$~~~~~~\times \left\{(x_{1}+ix_{2})y_{3}-
x_{3}(y_{1}+iy_{2})\right\}^{\lambda},~~~~
\lambda \leq q \leq \ell,~~~\lambda=0,1 . \eqno (14) $$ 

\noindent
$Q_{q}^{\ell \lambda}({\bf x,y})$ is the common eigenfunction
of ${\bf L}^{2}$, $L_{3}$, $\bigtriangleup_{x}$, 
$\bigtriangleup_{y}$, $\bigtriangleup_{xy}$, and the parity 
with the eigenvalues $\ell(\ell+1)$, $\ell$, $0$, $0$, $0$, 
and $(-1)^{\ell+\lambda}$, respectively, 
where ${\bf L}^{2}$, $L_{3}$ are the total angular 
momentum operators [see Eq. (9)], $\bigtriangleup_{x}$ 
and $\bigtriangleup_{y}$ are the Laplace operators respectively 
with respect to the position vectors ${\bf x}$ and ${\bf y}$ 
[see Eq. (3)], and $\bigtriangleup_{xy}$ is defined as
$$\bigtriangleup_{xy}=\displaystyle {\partial^{2} \over 
\partial x_{1}\partial y_{1} }+ \displaystyle {\partial^{2} \over 
\partial x_{2}\partial y_{2} } +\displaystyle {\partial^{2} \over 
\partial x_{3}\partial y_{3} }. \eqno (15) $$

\noindent
Their partners with the smaller eigenvalues of $L_{3}$ can 
be calculated from them by the lowering operator 
$L_{-}$ (see Eq. (9) and [20]). 

Now, the solutions to the Schr\"{o}dinger equation (1),
which are the common eigenfunctions of ${\bf L}^{2}$, $L_{3}$ 
and the parity with the eigenvalues $\ell(\ell+1)$, $\ell$,
and $(-1)^{\ell+\lambda}$, respectively, are generally written as
$$\Psi_{\ell \lambda}({\bf x,y})=\displaystyle \sum_{q=\lambda}^{\ell}~
\psi^{\ell \lambda}_{q}Q_{q}^{\ell \lambda}({\bf x,y}),~~~~~\lambda=0,1,
\eqno (16) $$

\noindent
where $\psi^{\ell \lambda}_{q}$ are the functions of the radial 
variables $\xi_{j}$. Due to conservation of the angular momentum and 
parity, $\Psi_{\ell \lambda}({\bf x,y})$ with different subscripts
$\ell$ and $\lambda$ are separated in the Schr\"{o}dinger equation. 
Recall that for $S$ wave ($\ell=0$), $Q_{0}^{00}({\bf x,y})=1$, and
$\Psi_{0 0}({\bf x,y})=\psi_{0}^{00}$. The 
wavefunction for $S$ wave only depends upon the radial variables. 

Substituting the wavefunctions (16) into the Schr\"{o}dinger 
equation (1) and (9), we obtain the radial equations for the 
radial functions $\psi^{\ell \lambda}_{q}$:
$$\displaystyle -{\hbar^{2} \over 2M} \left\{ \bigtriangleup
\psi^{\ell \lambda}_{q} +4q \displaystyle 
{\partial \psi^{\ell \lambda}_{q} \over \partial \xi_{1}}
+4(\ell-q+\lambda) \displaystyle 
{\partial \psi^{\ell \lambda}_{q} \over \partial \xi_{2}}
+2(q-\lambda) \displaystyle 
{\partial \psi^{\ell \lambda}_{q-1} \over \partial \xi_{3}}
+2(\ell-q) \displaystyle 
{\partial \psi^{\ell \lambda}_{q+1} \over \partial \xi_{3}} \right\} $$
$$=(E-V) \psi^{\ell \lambda}_{q},~~~~~
\lambda\leq q \leq \ell,~~~~~\lambda =0,1. \eqno (17) $$

\section{RADIAL VARIABLES}

In order to separate the radial variables in the radial equations 
(17), we define new radial variables $\rho$, $\alpha$ and $\beta$
through the complex vector coordinates [19]:
$$4k^{2}\left({\bf x}+i{\bf y}\right)^{2}=-\rho^{2}e^{-i\beta}\sin \alpha, $$
$$\rho=2k\left\{ \xi_{1}+\xi_{2}\right\}^{1/2},~~~~~
\cos \alpha =2\left(\xi_{1}\xi_{2}-\xi_{3}^{2}
\right)^{1/2}\left(\xi_{1}+\xi_{2}\right)^{-1},  $$
$$\tan \beta=2\xi_{3}\left(\xi_{2}-\xi_{1} \right)^{-1} , ~~~~~
\sin \beta= 2\xi_{3} \left[\left(\xi_{2}-\xi_{1}\right)^{2}
+4\xi_{3}^{2}\right]^{-1/2 } ,   $$
$$k^{2}=-2ME/\hbar^{2},~~~~~0\leq \rho < \infty,~~~~~
0\leq \alpha \leq \pi/2,~~~~~ -\pi<\beta \leq \pi. \eqno (18) $$

\noindent
Through replacement of variables, the Laplace operator 
$\bigtriangleup$ in Eq. (9) becomes 
$$\bigtriangleup=\displaystyle {4k^{2}  \over \rho^{5}} 
\displaystyle {\partial \over \partial \rho} \rho^{5} \displaystyle 
 {\partial \over \partial \rho} + \displaystyle {16k^{2}  \over \rho^{2}} T,
\eqno (19) $$
$$T= \displaystyle {1 \over \sin( 2\alpha)}
\displaystyle {\partial \over \partial \alpha}\sin (2\alpha) \displaystyle 
{\partial \over \partial \alpha}
+\displaystyle {1 \over \sin^{2} \alpha} \displaystyle 
{\partial^{2} \over \partial \beta^{2}} . \eqno (20) $$

Denoting by $Z_{n,m}(\alpha, \beta)$ the eigenfunction of $T$,
and expressing it as
$$Z_{n,m}(\alpha, \beta)=e^{im\beta} \left(\sin \alpha \right)^{|m|}
F(\zeta),~~~~~\zeta=\sin^{2} \alpha,  $$

\noindent
we have 
$$TZ_{n,m}(\alpha, \beta)=e^{im\beta} \left(\sin \alpha \right)^{|m|}
~~~~~~~~~~~~~~~~~~~~~~~~~~~~~~~~~~~~~~~~~~~~~~~~~~~~~~~~~~~~~~~~~$$
$$\times~\left\{4\zeta(1-\zeta)\displaystyle {d^{2} \over d\zeta^{2} }
+4\left[(|m|+1)-(|m|+2)\zeta\right]\displaystyle {d \over d\zeta}
-|m|(|m|+2)\right\} F(\zeta) . $$

\noindent
Let $F(\zeta)$ be the hypergeometric function
$F(-n,~n+|m|+1,~|m|+1,~\zeta)$, satisfying:
$$\left\{\zeta(1-\zeta)\displaystyle {d^{2} \over d\zeta^{2} }
+\left[(|m|+1)-(|m|+2)\zeta\right]\displaystyle {d \over d\zeta}
\right\} F(-n,~n+|m|+1,~|m|+1,~\zeta)$$
$$=-n(n+|m|+1)F(-n,~n+|m|+1,~|m|+1,~\zeta) . $$

\noindent
Thus, 
$$TZ_{n,m}(\alpha, \beta)=-\Lambda_{nm}Z_{n,m}(\alpha, \beta),
~~~~~\Lambda_{nm}=(2n+|m|)(2n+|m|+2), $$
$$Z_{n,m}(\alpha,\beta)=\left(\displaystyle {2n+|m|+1 \over \pi}
\right)^{1/2}e^{im\beta}\displaystyle \sum_{r=0}^{n}~
\displaystyle {(-1)^{r} (n+|m|+r)!(\sin \alpha)^{2r+|m|}
\over r!(n-r)!(|m|+r)! }  \eqno (21) $$

\noindent
where the normalization factor is included in front of 
$Z_{n,m}(\alpha,\beta)$ so that it satisfies
$$\displaystyle \int_{-\pi}^{\pi} d\beta \int_{0}^{\pi/2}
d\alpha \sin \alpha \cos \alpha Z_{n,m}(\alpha,\beta)^{*}
Z_{n',m'}(\alpha,\beta)=\delta_{nn'}\delta_{mm'}.
\eqno (22) $$

\noindent
The orthogonal condition is obvious from Eq. (21), and the
normalization condition can be proved by the identity (A2) in Appendix:
$$\displaystyle \int_{-\pi}^{\pi} d\beta \int_{0}^{\pi/2}
d\alpha \sin \alpha \cos \alpha |Z_{n,m}(\alpha,\beta)|^{2} 
~~~~~~~~~~~~~~~~~~~~~~~~~~~~~~~~~~~~~~~~~~~~~~~~~~~~~~~~~$$
$$=(2n+|m|+1) \sum_{r=0}^{n}~
\displaystyle {(-1)^{r} (n+|m|+r)!\over r!(n-r)!(|m|+r)! }  
\displaystyle \sum_{s=0}^{n}~\displaystyle {(-1)^{s} (n+|m|+s)!
\over s!(n-s)!(|m|+s)!(r+s+|m|+1) }  $$
$$=(2n+|m|+1) \sum_{r=0}^{n}~
\displaystyle {(-1)^{r} (n+|m|+r)!\over r!(n-r)!(|m|+r)! }  
~~~~~~~~~~~~~~~~~~~~~~~~~~~~~~~~~~~~~~~~~~~~~~~~~$$
$$~~~~\times~\displaystyle {(-1)^{n}r(r-1)\cdots (r-n+1) \over (r+|m|+1)
(r+|m|+2)\cdots (r+|m|+n+1) }=1,  $$

\noindent
where only one term ($r=n$) in the summation is non-vanishing. 
As a matter of fact, $Z_{nm}(\alpha, \beta)$ can also be expressed
by the Wigner $D$-function [21,16]:
$$Z_{n,m}(\alpha,\beta)=\left(\displaystyle {2n+m+1 \over \pi}
\right)^{1/2}e^{im\beta} d^{n+m/2}_{(-m/2)(m/2)}(2\alpha),
~~~~~{\rm when}~~m\geq 0. \eqno (23) $$

The Coulomb potential (4) depends only upon the radial
variables. It is straightforward to show
$$|{\bf r}_{1} - {\bf r}_{2}|^{2}=\displaystyle {m_{1}+m_{2}
\over m_{1}m_{2} }~\displaystyle {\rho^{2} \over 8k^{2} }~
\left\{ 1-\sin \alpha \cos (\beta -\beta_{3}) \right\}, 
\eqno (24) $$

\noindent
and those formulas obtained by replacing the subscripts $(1,2,3)$ 
cyclically. $\beta_{3}$ is the radial variable $\beta$ when the 
first particle coincides with the second particle, and similar
for $\beta_{1}$ and $\beta_{2}$. The calculation results are
$$\beta_{1}=\pi,~~~~~\sin \beta_{2}=\displaystyle  
{ 2\sqrt{m_{1}m_{2}m_{3}} \over m_{3}+m_{1}m_{2}},~~~~~
\cos \beta_{2}= \displaystyle { m_{3}-m_{1}m_{2} \over 
m_{3}+m_{1}m_{2} } $$
$$\sin \beta_{3}=\displaystyle  
{- 2\sqrt{m_{1}m_{2}m_{3}} \over m_{2}+m_{1}m_{3}},~~~~~
\cos \beta_{3}= \displaystyle { m_{2}-m_{1}m_{3} \over 
m_{2}+m_{1}m_{3} }  \eqno (25) $$

\noindent
By the way, when any two particles coincide with each other, 
${\bf x} \parallel {\bf y}$ and $\alpha=\pi/2$. 

Now, taking the Fourier-series expansion for 
$(1-\sin \alpha \cos \beta)^{-1/2}$, we have
$$\left( 1-\sin \alpha \cos \beta \right)^{-1/2}
=\displaystyle \sum_{r=-\infty}^{\infty} T_{r}(\sin \alpha)e^{ir\beta},$$
$$T_{r}(\sin \alpha)=T_{-r}(\sin \alpha)=\displaystyle \sum_{t=0}^{\infty}~
\displaystyle {(4t+2r)! (\sin \alpha)^{2t+r} \over 
8^{2t+r}t!(t+r)!(2t+r)!} ,  \eqno (26) $$

Substituting Eqs. (24) and (26) into Eq. (4), we obtain
$$\displaystyle {\rho \over \sqrt{8}ke^{2}}V(\rho,\alpha,\beta)
Z_{n',m'}(\alpha, \beta)=\displaystyle \sum_{nm}~Z_{n,m}(\alpha, \beta)
C(|m'-m|)D(n,m,n',m'), \eqno (27)$$ 
$$D(n,m,n',m')=D(n,-m,n',-m')=D(n',m',n,m)
~~~~~~~~~~~~~~~~~~~~~~~~~~~~~~~~~$$
$$=\displaystyle \sum_{r=-\infty}^{\infty}
\int_{-\pi}^{\pi} d\beta 
\int_{0}^{\pi/2} d\alpha \sin \alpha \cos \alpha 
Z_{n,m}(\alpha, \beta)^{*} T_{r}(\sin \alpha)e^{ir\beta}
Z_{n',m'}(\alpha, \beta) ~~~~~~~~~~~$$
$$=\displaystyle \left[(2n+|m|+1)(2n'+|m'|+1)\right]^{1/2}
\displaystyle \sum_{t=0}^{\infty}~\displaystyle 
{8^{-2t-|m-m'|}(4t+2|m-m'|)! \over
t!(t+|m-m'|)!(2t+|m-m'|)! }~$$
$$\times~\displaystyle \sum_{r=0}^{n}~
\displaystyle {(-1)^{r}(n+|m|+r)! \over r!(n-r)!(|m|+r)!}
~\displaystyle \sum_{s=0}^{n'}~
\displaystyle {(-1)^{s}(n'+|m'|+s)! \over s!(n'-s)!(|m'|+s)!}$$
$$\times~
\left[t+r+s+1+(|m|+|m'|+|m-m'|)/2\right]^{-1} , \eqno (28) $$
$$C(n)=\sqrt{\displaystyle {m_{2}m_{3}
\over m_{2}+m_{3} }}~Z_{2}Z_{3}e^{-in\beta_{1}}+
\sqrt{\displaystyle {m_{3}m_{1}
\over m_{3}+m_{1} }}~Z_{3}Z_{1}e^{-in\beta_{2}}+
\sqrt{\displaystyle {m_{1}m_{2}
\over m_{1}+m_{2} }}~Z_{1}Z_{2}e^{-in\beta_{3}}.
\eqno (29) $$

\noindent
The formula for $D(n,m,n',m')$ contains double alternating series, 
which can be calculated analytically in terms of the identities,
given in Appendix. For definiteness, we assume $n\geq n'$, 
and first calculate the summation over $s$ by Eq. (A2). 
Otherwise, we exchange $(n,m)$ from $(n',m')$ in $D(n,m,n',m')$.
$$\displaystyle \sum_{s=0}^{n'}\displaystyle {(-1)^{s}(n'+|m'|+s)! 
\over s!(n'-s)!(|m'|+s)!}~
\left[t+r+s+1+(|m|+|m'|+|m-m'|)/2\right]^{-1} $$
$$=\displaystyle {(-1)^{n'}(|m|+b+r)(|m|+b+r+1)\cdots (|m|+b+r+n'-1)
\over (|m|+a+r)(|m|+a+r+1)\cdots (|m|+a+r+n')}, $$
$$a=t+1+(|m-m'|-|m|+|m'|)/2,~~~~~b=a-n'-|m'|. $$

\noindent
Then, in terms of the identities in Appendix, we are able to express
the summation over $r$ analytically.

The remaining differential operators in the radial equations (17)
can also be expressed by those with respect to the new radial variables:
$$\displaystyle {\partial \over \partial \xi_{1}}=
\displaystyle {2k^{2} \over \rho} ~
\displaystyle {\partial \over \partial \rho}
-\displaystyle {4k^{2} \over \rho^{2}} \left\{
\tan \alpha \displaystyle {\partial \over \partial \alpha}
+\left[\displaystyle {\cos \beta \over \cos \alpha} ~
\displaystyle {\partial \over \partial \alpha}
-\displaystyle {\sin \beta \over \sin \alpha} ~
\displaystyle {\partial \over \partial \beta} \right] \right\}, $$
$$\displaystyle {\partial \over \partial \xi_{2}}=
\displaystyle {2k^{2} \over \rho} ~
\displaystyle {\partial \over \partial \rho}
-\displaystyle {4k^{2} \over \rho^{2}} \left\{
\tan \alpha \displaystyle {\partial \over \partial \alpha}
-\left[\displaystyle {\cos \beta \over \cos \alpha} ~
\displaystyle {\partial \over \partial \alpha}
-\displaystyle {\sin \beta \over \sin \alpha} ~
\displaystyle {\partial \over \partial \beta} \right] \right\}, $$
$$\displaystyle {\partial \over \partial \xi_{3}}=
\displaystyle {8k^{2} \over \rho^{2}} \left\{
\displaystyle {\sin \beta \over \cos \alpha} ~
\displaystyle {\partial \over \partial \alpha}
+\displaystyle {\cos \beta \over \sin \alpha} ~
\displaystyle {\partial \over \partial \beta} \right\}.
\eqno (30) $$

Applying them to $Z_{n,m}(\alpha, \beta)$, we obtain
$$\tan \alpha \displaystyle {\partial \over \partial \alpha}
Z_{n,m}(\alpha, \beta)=(2n+|m|)Z_{n,m}(\alpha, \beta)~~~~~~~~~~
~~~~~~~~~~~~~~$$
$$+~\displaystyle \sum_{r=0}^{n-1}~2(-1)^{n+r}\left\{(2r+|m|+1)
(2n+|m|+1)\right\}^{1/2}Z_{rm}(\alpha, \beta),  \eqno (31) $$

\noindent
and when $m>0$, 
$$\left\{\displaystyle {\cos \beta \over \cos \alpha} ~
\displaystyle {\partial \over \partial \alpha}
-\displaystyle {\sin \beta \over \sin \alpha} ~
\displaystyle {\partial \over \partial \beta} \right\} 
Z_{n,\pm m}(\alpha, \beta) ~~~~~~~~~~~~~~~~~~~~~~~~~~~~~~~~~~~~~~$$
$$=\displaystyle \sum_{r=0}^{n-1}~(-1)^{n+r}\left\{(2r+m+2)
(2n+m+1)\right\}^{1/2}Z_{r,\pm (m+1)}(\alpha, \beta)$$
$$+\displaystyle \sum_{r=0}^{n}~(-1)^{n+r}\left\{(2r+m)
(2n+m+1)\right\}^{1/2}Z_{r,\pm (m-1)}(\alpha, \beta), $$
$$\left\{\displaystyle {\cos \beta \over \cos \alpha} ~
\displaystyle {\partial \over \partial \alpha}
-\displaystyle {\sin \beta \over \sin \alpha} ~
\displaystyle {\partial \over \partial \beta} \right\} 
Z_{n,0}(\alpha, \beta) ~~~~~~~~~~~~~~~~~~~~~~~~~~~~~~~~~~~~~~$$
$$=\displaystyle \sum_{r=0}^{n-1}~(-1)^{n+r}\left\{(2r+2)
(2n+1)\right\}^{1/2}\left\{Z_{r,1}(\alpha, \beta)~
+Z_{r,(-1)}(\alpha, \beta) \right\}, $$
$$\pm i\left\{\displaystyle {\sin \beta \over \cos \alpha} ~
\displaystyle {\partial \over \partial \alpha}
+\displaystyle {\cos \beta \over \sin \alpha} ~
\displaystyle {\partial \over \partial \beta} \right\} 
Z_{n,\pm m}(\alpha, \beta) ~~~~~~~~~~~~~~~~~~~~~~~~~~~~~~~~~~~~~~$$
$$=\displaystyle \sum_{r=0}^{n-1}~(-1)^{n+r}\left\{(2r+m+2)
(2n+m+1)\right\}^{1/2}Z_{r,\pm (m+1)}(\alpha, \beta) $$
$$-\displaystyle \sum_{r=0}^{n}~(-1)^{n+r}\left\{(2r+m)
(2n+m+1)\right\}^{1/2}Z_{r,\pm (m-1)}(\alpha, \beta), $$
$$i\left\{\displaystyle {\sin \beta \over \cos \alpha} ~
\displaystyle {\partial \over \partial \alpha}
+\displaystyle {\cos \beta \over \sin \alpha} ~
\displaystyle {\partial \over \partial \beta} \right\} 
Z_{n,0}(\alpha, \beta) ~~~~~~~~~~~~~~~~~~~~~~~~~~~~~~~~~~~~~~$$
$$=\displaystyle \sum_{r=0}^{n-1}~(-1)^{n+r}\left\{(2r+2)
(2n+1)\right\}^{1/2}\left\{Z_{r,1}(\alpha, \beta)~
-Z_{r,(-1)}(\alpha, \beta) \right\}.  \eqno (32) $$

Now, expanding the radial functions $\psi^{\ell \lambda}_{q}$ as
$$\psi^{\ell \lambda}_{q}(\rho, \alpha, \beta)=
\displaystyle \sum_{n=0}^{\infty} \displaystyle \sum_{m=-\infty}^{\infty}~
R^{\ell \lambda}_{q,n,m}(\rho)Z_{n,m}(\alpha, \beta),
\eqno (33) $$

\noindent
and substituting them into the radial equations (17), we obtain 
a set of coupled ordinary differential equations for the 
functions $R^{\ell \lambda}_{q,n,m}(\rho)$:
$$\left\{\displaystyle \rho^{2} {\partial^{2} \over \partial \rho^{2}}
+ \rho \left(5+2\ell+2\lambda \right)
\displaystyle {\partial \over \partial \rho}
-\displaystyle {\rho^{2} \over 4} 
-4[\Lambda_{nm}+(\ell+\lambda)(2n+m)] \right\}
R^{\ell \lambda}_{q,n,\pm m}(\rho) $$
$$-8(\ell+\lambda)\displaystyle \sum_{r=n+1}^{\infty}~(-1)^{n+r}
\left\{(2n+m+1)(2r+m+1)\right\}^{1/2}R^{\ell \lambda}_{q,r, \pm m}(\rho)$$
$$+4(\ell-2q+\lambda)\left\{
\displaystyle \sum_{r=n+1}^{\infty}~(-1)^{n+r}\left\{(2n+m+1)
(2r+m)\right\}^{1/2}R^{\ell \lambda}_{q,r,\pm (m-1)}(\rho) \right. $$
$$\left. +\displaystyle \sum_{r=n}^{\infty}~(-1)^{n+r}\left\{(2n+m+1)
(2r+m+2)\right\}^{1/2}R^{\ell \lambda}_{q,r,\pm (m+1)}(\rho) 
\right\}$$
$$\mp i4(q-\lambda)\left\{
\displaystyle \sum_{r=n+1}^{\infty}~(-1)^{n+r}\left\{(2n+m+1)
(2r+m)\right\}^{1/2}R^{\ell \lambda}_{(q-1),r,\pm (m-1)}(\rho)\right. $$
$$\left. -\displaystyle \sum_{r=n}^{\infty}~(-1)^{n+r}\left\{(2n+m+1)
(2r+m+2)\right\}^{1/2}R^{\ell \lambda}_{(q-1),r,\pm (m+1)}(\rho)\right\}$$
$$\mp i4(\ell-q)\left\{
\displaystyle \sum_{r=n+1}^{\infty}~(-1)^{n+r}\left\{(2n+m+1)
(2r+m)\right\}^{1/2}R^{\ell \lambda}_{(q+1),r,\pm (m-1)}(\rho)\right. $$
$$\left. -\displaystyle \sum_{r=n}^{\infty}~(-1)^{n+r}\left\{(2n+m+1)
(2r+m+2)\right\}^{1/2}R^{\ell \lambda}_{(q+1),r,\pm (m+1)}(\rho)\right\}$$
$$=\left(\displaystyle {\sqrt{2}Me^{2}\over \hbar^{2} k} \right)
\displaystyle \sum_{r=0}^{\infty}\sum_{m'=-\infty}^{\infty}~
C(|m' \mp m|)D(n,\pm m,r,m')\rho R^{\ell \lambda}_{q,r,m'}(\rho), $$ 
$$\left\{\displaystyle \rho^{2} {\partial^{2} \over \partial \rho^{2}}
+ \rho \left(5+2\ell+2\lambda \right)
\displaystyle {\partial \over \partial \rho}
-\displaystyle {\rho^{2} \over 4} 
-4[\Lambda_{n0}+2n(\ell+\lambda)] \right\}
R^{\ell \lambda}_{q,n,0}(\rho) $$
$$-8(\ell+\lambda)\displaystyle \sum_{r=n+1}^{\infty}~(-1)^{n+r}
\left\{(2n+1)(2r+1)\right\}^{1/2}R^{\ell \lambda}_{q,r,0}(\rho)$$
$$+4(\ell-2q+\lambda)\displaystyle \sum_{r=n}^{\infty}~(-1)^{n+r}
\left\{(2n+1)(2r+2)\right\}^{1/2}
\left\{R^{\ell \lambda}_{q,r,1}(\rho)+
R^{\ell \lambda}_{q,r,(-1)}(\rho) \right\}$$
$$+i4(q-\lambda)\displaystyle \sum_{r=n}^{\infty}~(-1)^{n+r}
\left\{(2n+1)(2r+2)\right\}^{1/2}\left\{R^{\ell \lambda}_{(q-1),r,1}(\rho)
-R^{\ell \lambda}_{(q-1),r,(-1)}(\rho)\right\}$$
$$+i4(\ell-q)\displaystyle \sum_{r=n}^{\infty}~(-1)^{n+r}
\left\{(2n+1)(2r+2)\right\}^{1/2}\left\{R^{\ell \lambda}_{(q+1),r,1}(\rho)
-R^{\ell \lambda}_{(q+1),r,(-1)}(\rho)\right\}$$
$$=\left(\displaystyle {\sqrt{2}Me^{2}\over \hbar^{2} k} \right)
\displaystyle \sum_{r=0}^{\infty}\sum_{m'=-\infty}^{\infty}~
C(|m'|)D(n,0,r,m')\rho R^{\ell \lambda}_{q,r,m'}(\rho), 
 \eqno (34) $$ 

\noindent
where $m> 0$.

Because the differential operators in the set of coupled ordinary 
differential equations (34) are same, it can be reduced to
a system of linear algebraic equations by the spectral
decomposition of $R^{\ell \lambda}_{q,n,m}(\rho)$ in terms of the
associated Laguerre polynomials:
$$R^{\ell \lambda}_{q,n,m}(\rho)=e^{-\rho/2}
\displaystyle \sum_{p=0}^{\infty}~f^{\ell \lambda}_{p,q,n,m}
L_{p}^{(2\ell+2\lambda+4)}(\rho), \eqno (35) $$

\noindent
where
$$L_{n}^{(m)}(\rho)=\displaystyle {\rho^{-m}e^{\rho} \over n!}
\displaystyle {d^{n} \over d\rho^{n}}\left(e^{-\rho}\rho^{n+m}\right)
=\displaystyle \sum_{r=0}^{n}~\displaystyle {(-1)^{n-r}(n+m)!\rho^{n-r}
\over r!(n-r)!(n-r+m)! }, 
~~~~~n\geq 0,  \eqno (36) $$

\noindent
satisfying
$$\rho \displaystyle {d^{2}\over d\rho^{2}} L_{n}^{(m)}(\rho)
+(m+1-\rho) \displaystyle {d \over d\rho} L_{n}^{(m)}(\rho)
+n L_{n}^{(m)}(\rho)=0, $$
$$\rho L_{n}^{(m)}(\rho)=-(n+1)L_{n+1}^{(m)}(\rho)+
(2n+m+1)L_{n}^{(m)}(\rho)-(n+m)L_{n-1}^{(m)}(\rho), $$
$$\displaystyle \int_{0}^{\infty}d\rho e^{-\rho}\rho^{m}
L_{n}^{(m)}(\rho)L_{n'}^{(m)}(\rho)=\delta_{nn'}
\displaystyle {(n+m)! \over n!}. \eqno (37) $$

\noindent
Thus,
$$\left\{\displaystyle \rho^{2} {\partial^{2} \over \partial \rho^{2}}
+ \rho \left(5+2\ell+2\lambda \right)
\displaystyle {\partial \over \partial \rho}
-\displaystyle {\rho^{2} \over 4}  \right\}
e^{-\rho/2}L_{p}^{(2\ell+2\lambda+4)}(\rho) $$
$$=-e^{-\rho/2}(p+\ell+\lambda+5/2)\rho L_{p}^{(2\ell+2\lambda+4)}(\rho).
\eqno (38) $$

In summary, the solution of the Schr\"{o}dinger equation
for the three-body problem with the given angular momentum
$\ell$ and parity $(-1)^{\ell+\lambda}$, where $\lambda=0$ 
or $1$, can be expressed as an infinite series:
$$\Psi_{\ell \lambda}({\bf x,y})=e^{-\rho/2}\displaystyle 
\sum_{q=\lambda}^{\ell}~
\displaystyle \sum_{p=0}^{N_{1}}
\displaystyle \sum_{n=0}^{N_{2}}~
\displaystyle \sum_{m=-N_{3}}^{N_{3}}
f^{\ell \lambda}_{p,q,n,m}
L_{p}^{(2\ell+2\lambda+4)}(\rho)Z_{n,m}(\alpha, \beta)
Q_{q}^{\ell \lambda}({\bf x,y}),
\eqno (39) $$

\noindent
where $N_{j}$ are infinity, and the coefficients 
$f^{\ell \lambda}_{p,q,n,m}$ satisfy
a system of linear algebraic equations:
$$\left(\displaystyle { \hbar^{2} k \over \sqrt{2}Me^{2}} \right)
\left\{(p+\ell+\lambda+3/2)p f^{\ell \lambda}_{p-1,q,n,\pm m}
-[4\Lambda_{nm}+4(\ell+\lambda)(2n+m) \right. $$
$$+2(p+\ell+\lambda+5/2)^{2}] f^{\ell \lambda}_{p,q,n, \pm m}
+(p+\ell+\lambda+7/2)(p+2\ell+2\lambda+5) f^{\ell \lambda}_{p+1,q,n,\pm m}$$
$$-8(\ell+\lambda)\displaystyle \sum_{r=n+1}^{N_{2}}~(-1)^{n+r}
\left[(2n+m+1)(2r+m+1)\right]^{1/2}
f^{\ell \lambda}_{p,q,r,\pm m}$$
$$+4(\ell-2q+\lambda)\left[
\displaystyle \sum_{r=n+1}^{N_{2}}~(-1)^{n+r}\left\{(2n+m+1)
(2r+m)\right\}^{1/2}f^{\ell \lambda}_{p,q,r,\pm (m-1)} \right. $$
$$\left. +\displaystyle \sum_{r=n}^{N_{2}}~(-1)^{n+r}\left\{(2n+m+1)
(2r+m+2)\right\}^{1/2}f^{\ell \lambda}_{p,q,r,\pm (m+1)}
\right]$$
$$\mp i4(q-\lambda)\left[
\displaystyle \sum_{r=n+1}^{N_{2}}~(-1)^{n+r}\left\{(2n+m+1)
(2r+m)\right\}^{1/2}f^{\ell \lambda}_{p,(q-1),r,\pm (m-1)}\right. $$
$$\left. -\displaystyle \sum_{r=n}^{N_{2}}~(-1)^{n+r}
\left\{(2n+m+1)(2r+m+2)\right\}^{1/2}
f^{\ell \lambda}_{p,(q-1),r,\pm (m+1)}\right]$$
$$\mp i4(\ell-q)\left[
\displaystyle \sum_{r=n+1}^{N_{2}}~(-1)^{n+r}\left\{(2n+m+1)
(2r+m)\right\}^{1/2}f^{\ell \lambda}_{p,(q+1),r,\pm (m-1)}\right. $$
$$\left.\left. -\displaystyle \sum_{r=n}^{N_{2}}~(-1)^{n+r}\left\{(2n+m+1)
(2r+m+2)\right\}^{1/2}f^{\ell \lambda}_{p,(q+1),r,\pm (m+1)}\right]
\right\} $$
$$=\displaystyle \sum_{r=0}^{N_{2}}\sum_{m'=-N_{3}}^{N_{3}}~
C(|m' \mp m|)D(n,\pm m,r,m')\left\{-p f^{\ell \lambda}_{(p-1),q,r,m'}
\right. $$
$$\left.+(2p+2\ell+2\lambda+5) f^{\ell \lambda}_{p,q,r,m'}
-(p+2\ell+2\lambda+5) f^{\ell \lambda}_{(p+1),q,r,m'} \right\},  $$ 
$$\left(\displaystyle { \hbar^{2} k \over \sqrt{2}Me^{2}} \right)
\left\{(p+\ell+\lambda+3/2)p f^{\ell \lambda}_{p-1,q,n,0}
-[4\Lambda_{n0}+8n(\ell+\lambda) \right. $$
$$+2(p+\ell+\lambda+5/2)^{2}] f^{\ell \lambda}_{p,q,n, 0}
+(p+\ell+\lambda+7/2)(p+2\ell+2\lambda+5) f^{\ell \lambda}_{p+1,q,n,0}$$
$$-8(\ell+\lambda)\displaystyle \sum_{r=n+1}^{N_{2}}~(-1)^{n+r}
\left[(2n+1)(2r+1)\right]^{1/2}
f^{\ell \lambda}_{p,q,r,0}$$
$$+4(\ell-2q+\lambda)
\displaystyle \sum_{r=n}^{N_{2}}~(-1)^{n+r}\left\{(2n+1)
(2r+2)\right\}^{1/2}\left(f^{\ell \lambda}_{p,q,r,1}
+f^{\ell \lambda}_{p,q,r,(-1)}\right)$$
$$+i4(q-\lambda)\displaystyle \sum_{r=n}^{N_{2}}~(-1)^{n+r}
\left\{(2n+1)(2r+2)\right\}^{1/2}
\left(f^{\ell \lambda}_{p,(q-1),r,1}-
f^{\ell \lambda}_{p,(q-1),r,(-1)}\right)$$
$$\left.+i4(\ell-q)\displaystyle \sum_{r=n}^{N_{2}}~(-1)^{n+r}\left\{(2n+1)
(2r+2)\right\}^{1/2}\left(f^{\ell \lambda}_{p,(q+1),r,1}
-f^{\ell \lambda}_{p,(q+1),r,(-1)} \right)\right\} $$
$$=\displaystyle \sum_{r=0}^{N_{2}}\sum_{m'=-N_{3}}^{N_{3}}~
C(|m'|)D(n,0,r,|m'|)\left\{-p f^{\ell \lambda}_{(p-1),q,r,m'}
\right. $$
$$\left.+(2p+2\ell+2\lambda+5) f^{\ell \lambda}_{p,q,r,m'}
-(p+2\ell+2\lambda+5) f^{\ell \lambda}_{(p+1),q,r,m'} \right\},
 \eqno (40) $$ 

\noindent
where $m>0$. The coefficients $C(n)$ and $D(n,m,n',m')$ were given 
in Eqs. (28) and (29). The square of the factor $\hbar^{2} k/(\sqrt{2}Me^{2})$ 
is the energy $-E$ in the unit $Me^{4}/\hbar^{2}$. This factor can be 
calculated by the condition that the coefficient determinant
of the coupled linear algebraic equations vanishes. This problem of
calculation can be changed to an eigenvalue problem. In the real 
calculation, the infinite series is truncated at integers $N_{j}$ 
large enough to make the coefficients at the truncated terms smaller 
than the permitted error.  

\section{THE HELIUM ATOM}

The problem of a helium atom is a typical three-body problem
with two identical particles. For this 
problem, we enumerate the helium nucleus to be the first 
particle, and two identical electrons to be the second and 
the third particles. Denote by $M_{He}$ and $M_{e}$ the
masses of the nucleus and the electron, whose electric charges
are $2e$ and $-e$, respectively. Thus, 
$$m_{1}=1-2m_{e},~~~~~m_{2}=m_{3}=m_{e},$$
$$Z_{1}=2,~~~~~~~~ Z_{2}=Z_{3}=-1, $$
$${\bf x}=-\sqrt{(1-2m_{e})/(2m_{e})} ~{\bf r_{1}},~~~~~
{\bf y}=(m_{e}/2)^{1/2}\left({\bf r_{2}-r_{3}}\right),$$
$$\beta_{1}=\pi,~~~~~~~~\beta_{2}=-\beta_{3}, \eqno (41) $$

\noindent
Therefore, the coefficient $C(n)$ in Eq. (29) becomes real. 
Hereafter we only remain the leading terms and the 
next leading terms with respect to the mass ratio $m_{e}=M_{e}/M$,
where $M=M_{He}+2M_{e}$ is the total mass of the helium atom.
$$\displaystyle {C(n) \over \sqrt{m_{e}}}\sim \left\{\begin{array}{ll}
\sqrt{1/2}-4(-1)^{n/2}(1-m_{e}/2)~~~~~
&{\rm when}~~n~~{\rm is~even}\\
-\sqrt{1/2}-4n(-1)^{(n-1)/2}m_{e} &{\rm when}~~n~~{\rm is~odd}, 
\end{array} \right. \eqno (42) $$

Although the spin-dependent interactions are neglected in
the Schr\"{o}dinger equation (1) for the helium atom, the
symmetric effects of the spins of two electrons have to be
taken account. The helium is called parahelium if the spinor
part of wavefunction is antisymmetric ($S=0$), and it is 
called orthohelium if the spinor wavefunction is symmetric 
($S=1$). Therefore, the spatial wavefunction of parahelium
is symmetric with respect to the permutation of two electrons,
and that of orthohelium is antisymmetric. 

On the other hand, in our notation, in the permutation of 
two electrons, ${\bf y}$ changes its sign and ${\bf x}$ remains 
invariant, namely, $\beta$ changes its sign, $\rho$ and $\alpha$ 
remain invariant, so that $L_{n}^{(m)}(\rho)$ remain invariant, 
$Z_{nm}(\alpha,\beta)$ changes to its complex conjugate, and 
$Q_{q}^{\ell \lambda}({\bf x,y})$ changes a factor $(-1)^{\ell-q+\lambda}$.
This symmetry gives some relations between the coefficients
$f^{\ell \lambda}_{p,q,n,m}$ in Eq. (39) such that the radial
function $\psi^{\ell \lambda}_{q}(\rho,\alpha,\beta)$ becomes real.

\subsection{$S$ wave in parahelium}

For $S$ wave in parahelium, the wavefunction only depends 
upon the radial variables, and is symmetric with respect to 
the permutation of two electrons, namely, 
$f^{00}_{p,0,n,-m}=f^{00}_{p,0,n,m}$. For simplicity, we briefly
denote the coefficient $f^{00}_{p,0,n,m}$ by $f_{p,n,m}$:
$$\Psi_{00}({\bf x,y})=e^{-\rho/2}
\displaystyle \sum_{p=0}^{N_{1}}
\displaystyle \sum_{n=0}^{N_{2}}~
\displaystyle \sum_{m=0}^{N_{3}}
f_{p,n,m}L_{p}^{(4)}(\rho)
\left(2-\delta_{m0}\right)\Re \left[Z_{n,m}(\alpha, \beta)\right].
\eqno (43) $$

\noindent
where $\Re [Z]$ denotes the real part of $Z$, and $\delta_{m0}$
is the Kronecker $\delta$ function. From Eq. (40),
the coefficients $f_{p,n,m}$ satisfy a real 
system of linear algebraic equations with the equation number 
$(N_{1}+1)(N_{2}+1)(N_{3}+1)$:
$$\left(\displaystyle { \hbar^{2} k \over \sqrt{2m_{e}}Me^{2}} \right)
\left\{\left(p+\displaystyle {3 \over 2}\right)p f_{p-1,n,m}
-\left[4\Lambda_{nm}+2\left(p+\displaystyle {5 \over 2}\right)^{2}
\right] f_{p,n,m} \right. $$
$$\left.~~~~~+\left(p+\displaystyle {7 \over 2}\right)(p+5) f_{p+1,n,m}
\right\} $$
$$~~~~=\displaystyle \sum_{r=0}^{N_{2}}
\sum_{m'=0}^{N_{3}}~\displaystyle {2-\delta_{m'0} \over 2\sqrt{m_{e}}}
\left[C(|m'- m|)D(n,m,r,m')+C(m'+ m)D(n,m,r,-m')\right] $$
$$~~~~~\times~\left\{-p f_{(p-1),r,m'}+(2p+5) f_{p,r,m'}
-(p+5) f_{(p+1),r,m'} \right\}.
 \eqno (44)  $$ 

\noindent
Therefore, the wavefunction $\Psi_{00}({\bf x,y})$ is real. 

The square of the factor $\hbar^{2} k/(\sqrt{2m_{e}}Me^{2})$ 
in Eq. (44) is the binding energy $-E$ in the unit $e^{2}/a_{0}$, 
where $a_{0}=\hbar^{2}/(M_{e}e^{2})$ is the Bohr radius. This factor 
can be calculated from the condition that the coefficient
determinant of Eq. (44) vanishes. The coefficient matrix on the
left hand side of Eq. (44) is a direct product of a 
$(N_{1}+1)$-dimensional matrix and a unit matrix. Removing it to 
the right hand side by right-multiplying with its inverse matrix, we
reduce the problem of calculating energy into an eigenvalue
problem. In the numerical calculation, we take $N_{1}=7$, $N_{2}=8$ 
and $N_{3}=16$, and obtain the energy for the ground state of a 
helium atom:
$$-E^{para}_{00}=2.88935\left(e^{2}/a_{0}\right)=78.62 {\rm eV}, 
\eqno (45) $$

\noindent
where the physical constants $e$, $\hbar$, $c$, $M_{e}$, and $M_{He}$
are quoted from the particle physics booklet, 1998. The experimental 
value [2] for the minimum energy required to remove both electrons 
from a helium atom is $2.90351e^{2}/a_{0}\sim 79.00$eV. 
The relative deviation of the calculated energy  
from the observed value is in the thousandths.

We also obtain the eigenfunction for the ground state of helium,
expressed as the series (43), which is normalized such that the 
largest coefficient in modulus is one ($f_{0,0,0}$). Those 
coefficients whose absolute values are larger than 0.001
are listed as follows:
$$f_{0,0,0}=1.0000,~~f_{0,0,1}=0.0411  ,~~f_{0,0,2}=-0.1047,~~
f_{0,0,4}=0.0209,~~f_{0,0,6}=-0.0076, $$
$$f_{0,0,8}=0.0030,~~f_{0,0,10}=-0.0016,~~f_{0,1,0}=-0.0899,~~
f_{0,1,1}=-0.0032,~~f_{0,1,2}=0.0227, $$
$$f_{0,1,4}=-0.0069,~~f_{0,1,6}=0.0032,~~f_{0,1,8}=-0.0015,~~
f_{0,2,0}=0.0204,~~  f_{0,2,2}=-0.0075,  $$
$$f_{0,2,4}=0.0029,~~f_{0,2,6}=-0.0016,~~f_{0,3,0}=-0.0069,~~
f_{0,3,2}=0.0032,~~  f_{0,3,4}=-0.0015, $$
$$f_{0,4,0}=0.0029 ,~~f_{0,4,2}=-0.0016 ,~~f_{0,5,0}=-0.0015,~~
f_{1,0,0}=-0.0440,~~ f_{1,0,1}=-0.0058, $$
$$f_{1,0,2}=0.0293,~~f_{1,0,4}=-0.0082,~~f_{1,0,6}=0.0031,~~
f_{1,0,8}=-0.0014,~~ f_{1,1,0}=0.0271, $$
$$f_{1,1,2}=-0.0085,~~f_{1,1,4}=0.0030,~~f_{1,1,6}=-0.0014,~~
f_{1,2,0}=-0.0081,~~f_{1,2,2}=0.0031,  $$
$$f_{1,2,4}=-0.0014,~~f_{1,3,0}=0.0030,~~f_{1,3,2}=-0.0014,~~
f_{1,4,0}=-0.0013,~~f_{2,0,0}=0.0036,  $$
$$f_{2,0,2}=-0.0037,~~f_{2,0,4}=0.0020 ,~~f_{2,1,0}=-0.0039,~~
f_{2,1,2}=0.0019,~~ f_{2,2,0}=0.0019. \eqno (46)  $$
\noindent
The remaining coefficients can be obtained from us upon request.

Changing the electric charge of the nucleus, we can obtain
the energies of the ground states of the helium-like ions. The 
calculation results and the observed results are listed in
Table 1.

\begin{center}

{\bf Table 1} $~~$ Calculated and observed energies of ground states

 for helium-like ions (in the unit $e^{2}/a_{0}$)

\vspace{3mm}
\begin{tabular}{c|c|c|c} \hline \hline
Ion & Calculated $^{a}$ 
& Observed $^{b}$
& Relative error $^{c}$ \\ \hline
H$^{-}$ & 0.520618 & & \\
He  & 2.88935 & 2.90351$\pm$0.00004 & 0.005\\
Li$^{+}$ & 7.25064 & 7.27980$\pm$0.00050 & 0.004\\
Be$^{++}$ & 13.6055 & 13.65600$\pm$0.00100 &0.004\\
B$^{+++}$ & 21.9543 & 22.03200$\pm$0.00150 &0.004 \\
C$^{++++}$ & 32.2973 & 32.40700$\pm$0.00400 & 0.003 \\ \hline
\end{tabular} 

\vspace{2mm}
\hspace{10mm}\parbox[t]{12cm}{a. By Eq. (43) with $N_{1}=7$, $N_{2}=8$ and 
$N_{3}=17$.\\  
b. See Refs. [22] and [23] \\ 
c. $(E_{obs}-E_{cal})/E_{obs}$}

\end{center}

\subsection{$S$ wave in orthohelium}

For $S$ wave in orthohelium, the wavefunction also depends 
upon the radial variables, but is antisymmetric with respect to 
the permutation of two electrons, namely, 
$f^{00}_{p,0,n,-m}=-f^{00}_{p,0,n,m}$. For simplicity, we briefly
denote the coefficient $f^{00}_{p,0,n,m}$ by $-ig_{p,n,m}$:
$$\Psi_{00}({\bf x,y})=2e^{-\rho/2}
\displaystyle \sum_{p=0}^{N_{1}}
\displaystyle \sum_{n=0}^{N_{2}}~
\displaystyle \sum_{m=1}^{N_{3}}
g_{p,n,m}L_{p}^{(4)}(\rho)
\Im \left[Z_{n,m}(\alpha, \beta)\right],
\eqno (47) $$

\noindent
where $\Im [Z]$ denotes the imaginary part of $Z$. From Eq. (40),
the coefficients $g_{p,n,m}$ satisfy a real 
system of linear algebraic equations with the equation number 
$(N_{1}+1)(N_{2}+1)N_{3}$:
$$\left(\displaystyle { \hbar^{2} k \over \sqrt{2m_{e}}Me^{2}} \right)
\left\{\left(p+\displaystyle {3 \over 2}\right)p g_{p-1,n,m}
-\left[4\Lambda_{nm}+2\left(p+\displaystyle {5 \over 2}\right)^{2}
\right] g_{p,n,m} \right.$$
$$\left.~~~~~+\left(p+\displaystyle {7 \over 2}\right)(p+5) g_{p+1,n,m}
\right\}$$
$$~~~~=\displaystyle \sum_{r=0}^{N_{2}}
\sum_{m'=1}^{N_{3}}~
\left[C(|m'- m|)D(n,m,r,m')-C(m'+ m)D(n,m,r,-m')\right]/\sqrt{m_{e}} $$
$$~~~~~\times~\left\{-p g_{(p-1),r,m'}+(2p+5) g_{p,r,m'}
-(p+5) g_{(p+1),r,m'} \right\}. \eqno (48)  $$ 

\noindent
Therefore, the wavefunction $\Psi_{00}({\bf x,y})$ is also real. 
Similarly, the energy of the ground state of the orthohelium
can be calculated as an eigenvalue problem. In the numerical calculation, 
we take $N_{1}=7$, $N_{2}=8$ and $N_{3}=17$, and obtain the energy 
for the ground state of the orthohelium:
$$-E_{00}^{ortho}=2.08039\left(e^{2}/a_{0}\right)=56.61 {\rm eV}. 
\eqno (49) $$

\noindent
The experimental value for it is 
$-E_{00}^{obs}=2.17524 \left(e^{2}/a_{0}\right)$=59.19eV. 
The relative error is 4\%. We will discuss this problem in Sec. VI.

\subsection{$P$ wave in parahelium}

There are two sets of solutions for the $P$ wave with the different
parities. The eigenfunctions $Q^{\ell \lambda}_{q}$ of the angular 
momentum [see Eq. (14)] are written explicitly as follows, where 
$q=0$ or $1$ when $\lambda=0$, and $q=1$ when $\lambda=1$:
$$Q^{10}_{1}({\bf x,y})=x_{1}+ix_{2},~~~~~
Q^{10}_{0}({\bf x,y})=y_{1}+iy_{2}, $$
$$Q^{11}_{1}({\bf x,y})=\left(x_{1}+ix_{2}\right)y_{3}
-x_{3}\left(y_{1}+iy_{2}\right). \eqno (50) $$

\noindent
From the condition that the spatial wavefunction in parahelium is 
symmetric with respect to the permutation of two electrons, 
we have 
$$f_{p,n,m}\equiv f^{10}_{p,1,n,m}=f^{10}_{p,1,n,-m},$$
$$-ig_{p,n,m}\equiv f^{10}_{p,0,n,m}=-f^{10}_{p,0,n,-m},$$
$$-ih_{p,n,m}\equiv f^{11}_{p,1,n,m}=-f^{11}_{p,1,n,-m}. \eqno (51) $$

\noindent
Thus, the wavefunctions are
$$\Psi_{10}({\bf x,y})=e^{-\rho/2}
\displaystyle \sum_{p=0}^{N_{1}}
\displaystyle \sum_{n=0}^{N_{2}}~
\displaystyle \sum_{m=0}^{N_{3}}
f_{p,n,m}L_{p}^{(6)}(\rho)
\left(2-\delta_{m0}\right)\Re \left[Z_{n,m}(\alpha, \beta)\right]
Q_{1}^{10}({\bf x,y}) $$
$$~~~~+2e^{-\rho/2}\displaystyle \sum_{p=0}^{N_{4}}
\displaystyle \sum_{n=0}^{N_{5}}~
\displaystyle \sum_{m=1}^{N_{6}}
g_{p,n,m}L_{p}^{(6)}(\rho)
\Im \left[Z_{n,m}(\alpha, \beta)\right]Q_{0}^{10}({\bf x,y}), $$
$$\Psi_{11}({\bf x,y})=2e^{-\rho/2}\displaystyle 
\displaystyle \sum_{p=0}^{N_{7}}
\displaystyle \sum_{n=0}^{N_{8}}~
\displaystyle \sum_{m=1}^{N_{9}}
h_{p,n,m}L_{p}^{(8)}(\rho)
\Im \left[Z_{n,m}(\alpha, \beta) \right]Q_{1}^{11}({\bf x,y}).
\eqno (52) $$

\noindent
From Eq. (40), the coefficients satisfy a real system of 
linear algebraic equations, so that the coefficients are real:
$$\left(\displaystyle { \hbar^{2} k \over \sqrt{2m_{e}}Me^{2}} \right)
\left\{\left(p+\displaystyle {5 \over 2}\right)p f_{p-1,n,m}
-\left[4\Lambda_{nm}+4(2n+m) +2\left(p+\displaystyle 
{7 \over 2}\right)^{2}\right] f_{p,n,m}\right. $$
$$~~~~~+\left(p+\displaystyle {9 \over 2}\right)(p+7) f_{p+1,n,m}
-8\displaystyle \sum_{r=n+1}^{N_{2}}~(-1)^{n+r}
\left[(2n+m+1)(2r+m+1)\right]^{1/2} f_{p,r,m}$$
$$~~~~~-4\left[\left(1-\delta_{m0}\right)\displaystyle 
\sum_{r=n+1}^{N_{2}}~(-1)^{n+r}\left\{(2n+m+1)
(2r+m)\right\}^{1/2}f_{p,r,m-1} \right. $$
$$~~~~~\left. +\left(1+\delta_{m0}\right)\displaystyle 
\sum_{r=n}^{N_{2}}~(-1)^{n+r}\left\{(2n+m+1)
(2r+m+2)\right\}^{1/2}f_{p,r,m+1} \right]$$
$$~~~~~-4\left[\left(1-\delta_{m0}\right)\displaystyle 
\sum_{r=n+1}^{N_{5}}~(-1)^{n+r}\left\{(2n+m+1)
(2r+m)\right\}^{1/2}g_{p,r,m-1}\right. $$
$$~~~~~\left. \left. -\left(1+\delta_{m0}\right)
\displaystyle \sum_{r=n}^{N_{5}}~(-1)^{n+r}
\left\{(2n+m+1)(2r+m+2)\right\}^{1/2}
g_{p,r,m+1}\right] \right\}$$
$$~~~=\displaystyle \sum_{r=0}^{N_{2}}\sum_{m'=0}^{N_{3}}~
\displaystyle {2-\delta_{m'0}\over 2\sqrt{m_{e}} }
\left\{C(|m'- m|)D(n,m,r,m')+C(m'+ m)D(n,m,r,-m')\right\} $$
$$~~~~~\times\left\{-p f_{p-1,r,m'}
+(2p+7) f_{p,r,m'}
-(p+7) f_{p+1,r,m'} \right\},  $$ 
$$\left(\displaystyle { \hbar^{2} k \over \sqrt{2m_{e}}Me^{2}} \right)
\left\{\left(p+\displaystyle {5 \over 2}\right)p g_{p-1,n, m}
-\left[4\Lambda_{nm}+4(2n+m)+2\left(p+\displaystyle 
{7 \over 2}\right)^{2}\right] g_{p,n, m} \right. $$
$$~~~~~+\left(p+\displaystyle {9 \over 2}\right)(p+7) g_{p+1,n, m}
-8\displaystyle \sum_{r=n+1}^{N_{5}}~(-1)^{n+r}
\left[(2n+m+1)(2r+m+1)\right]^{1/2} g_{p,r, m}$$
$$~~~~~+4\left[\displaystyle \sum_{r=n+1}^{N_{5}}~(-1)^{n+r}\left\{(2n+m+1)
(2r+m)\right\}^{1/2}g_{p,r,m-1} \right. $$
$$~~~~~\left. +\displaystyle \sum_{r=n}^{N_{5}}~(-1)^{n+r}\left\{(2n+m+1)
(2r+m+2)\right\}^{1/2}g_{p,r,m+1}\right] $$
$$~~~~~+4\left[\displaystyle \sum_{r=n+1}^{N_{2}}~(-1)^{n+r}\left\{(2n+m+1)
(2r+m)\right\}^{1/2}f_{p,r,m-1}\right. $$
$$~~~~~\left.\left. -\displaystyle \sum_{r=n}^{N_{2}}~(-1)^{n+r}\left\{(2n+m+1)
(2r+m+2)\right\}^{1/2}f_{p,r,m+1}\right] \right\} $$
$$~~~=\displaystyle \sum_{r=0}^{N_{5}}\sum_{m'=1}^{N_{6}}~
\left\{C(|m' -m|)D(n,m,r,m')-C(m'+ m)D(n,m,r,-m')\right\}/\sqrt{m_{e}}$$
$$~~~~~\times \left\{-p g_{p-1,r,m'}+(2p+7) g_{p,r,m'}
-(p+7) g_{(p+1),r,m'} \right\}, \eqno (53) $$ 
$$\left(\displaystyle { \hbar^{2} k \over \sqrt{2m_{e}}Me^{2}} \right)
\left\{\left(p+\displaystyle {7 \over 2}\right)p h_{p-1,n, m}
-\left[4\Lambda_{nm}+8(2n+m) 
+2\left(p+\displaystyle {9 \over 2}\right)^{2}\right] h_{p,n, m} \right. $$
$$~~~+\left(p+\displaystyle {11 \over 2}\right)(p+9) h_{p+1,n,m}
\left.-16\displaystyle \sum_{r=n+1}^{N_{8}}~(-1)^{n+r}
\left[(2n+m+1)(2r+m+1)\right]^{1/2}
h_{p,r,m} \right\}$$
$$~~=\displaystyle \sum_{r=0}^{N_{8}}\sum_{m'=1}^{N_{9}}~
\left\{C(|m'-m|)D(n,m,r,m')-C(m'+m)D(n,m,r,-m')\right\}/\sqrt{m_{e}}$$
$$~~~\times\left\{-p h_{p-1,r,m'}+(2p+9) h_{p,r,m'}
-(p+9) h_{p+1,r,m'} \right\}, \eqno (54) $$ 

\noindent
The energy can be calculated as an eigenvalue problem.
In the numerical calculation, we take $N_{1}=N_{4}=N_{7}=7$, 
$N_{2}=N_{5}=N_{8}=8$, $N_{3}=N_{6}=N_{9}=17$, and 
obtain the lowest energies of the $P$ wave with the odd parity 
and the even parity for the parahelium, respectively:
$$-E^{para}_{10}=2.02095\left(e^{2}/a_{0}\right)=54.99 {\rm eV}, 
\eqno (58) $$
$$-E^{para}_{11}=0.581291\left(e^{2}/a_{0}\right)=15.82 {\rm eV}. 
\eqno (59) $$

In the shell model of the atomic physics, the electrons in a 
helium atom are supposed to fill in the energy levels of a 
hydrogen-like atom with the electric charge $+2e$ according 
to the exclusion principle. In this model, the lowest-energy 
state of the $P$ wave with the odd parity in parahelium is
explained as the compound state of one $1S$ electron and one $2P$ 
electron. The observed energy is $-E_{10}^{obs}=2.1238(e^{2}/a_{0}$)
=57.79eV. The relative error for the calculated energy $-E^{para}_{10}$ 
is 5 \%. 

On the other hand, it is worthy to pay more attention to the 
existence of the lowest-energy state of the $P$ wave with the 
even parity in the parahelium. This state is forbidden for 
the electric dipole transition to the ground state of the 
parahelium, but it is allowed to the $P$ wave state with the 
odd parity in the parahelium. The energy difference between 
two $P$ wave states with different parities is
$$\Delta E=1.43966\left(e^{2}/a_{0}\right)=39.18{\rm eV}
=315969 cm^{-1}. \eqno (60) $$

\noindent
It is larger than the ionization energy (24.58eV) of the 
helium atom, which is the minimum energy required 
to remove one electron from the ground state of a helium atom.
In the shell model, this $P$-wave state has to be composed of two 
$P$-wave electrons. According to the Clebsch-Gordan coefficients 
[20], the angular part of its wavefunction is antisymmetric [see Eq. 
(50)]. Therefore, this state should be explained in the shell 
model as the compound state of one $2P$ and one $3P$ electrons, 
because its radial function has to be antisymmetric. However, 
the binding energy $-E$ of the compound state seems 
to be smaller than our calculated value. We will further discuss 
this state in Sec. VI.

\subsection{$P$ wave in orthohelium}

We sketch the calculation on the lowest-energy states of
the $P$ wave in orthohelium. From the condition that the 
spatial wavefunction in the orthohelium is antisymmetric 
with respect to the permutation of two electrons, 
we have 
$$-if_{p,n,m}\equiv f^{10}_{p,1,n,m}=-f^{10}_{p,1,n,-m},$$
$$g_{p,n,m}\equiv f^{10}_{p,0,n,m}=f^{10}_{p,0,n,-m},$$
$$h_{p,n,m}\equiv f^{11}_{p,1,n,m}=f^{11}_{p,1,n,-m}. \eqno (61) $$

\noindent
Thus, the wavefunctions are
$$\Psi_{10}({\bf x,y})=2e^{-\rho/2}
\displaystyle \sum_{p=0}^{N_{1}}
\displaystyle \sum_{n=0}^{N_{2}}~
\displaystyle \sum_{m=1}^{N_{3}}
f_{p,n,m}L_{p}^{(6)}(\rho)\Im \left[Z_{n,m}(\alpha, \beta)\right]
Q_{1}^{10}({\bf x,y}) $$
$$~~~~+e^{-\rho/2}\displaystyle \sum_{p=0}^{N_{4}}
\displaystyle \sum_{n=0}^{N_{5}}~
\displaystyle \sum_{m=0}^{N_{6}}
g_{p,n,m}L_{p}^{(6)}(\rho)\left(2-\delta_{m0}\right)
\Re \left[Z_{n,m}(\alpha, \beta)\right]Q_{0}^{10}({\bf x,y}), $$
$$\Psi_{11}({\bf x,y})=e^{-\rho/2}\displaystyle 
\displaystyle \sum_{p=0}^{N_{7}}
\displaystyle \sum_{n=0}^{N_{8}}~
\displaystyle \sum_{m=0}^{N_{9}}
h_{p,n,m}L_{p}^{(8)}(\rho)\left(2-\delta_{m0}\right)
\Re \left[Z_{n,m}(\alpha, \beta) \right]Q_{1}^{11}({\bf x,y}).
\eqno (62) $$

\noindent
From Eq. (40), the coefficients satisfy the real systems of 
linear algebraic equations, so that the coefficients are real.
The real systems of linear algebraic equations are similar to that
for the parahelium, except for changing some signs. We will not 
list those equations here, but list the calculated results for 
the energy, where the series are truncated at $N_{1}=N_{4}=N_{7}=7$, 
$N_{2}=N_{5}=N_{8}=8$, $N_{3}=N_{6}=N_{9}=17$, 
$$-E^{ortho}_{10}=2.04388\left(e^{2}/a_{0}\right)=55.62 {\rm eV}, 
\eqno (63) $$
$$-E^{ortho}_{11}=0.710413\left(e^{2}/a_{0}\right)=19.33 {\rm eV}. 
\eqno (64) $$

In the shell model of the atomic physics, the lowest-energy 
state of the $P$ wave with the odd parity in orthohelium is
also explained as the compound state of one $1S$ electron and one $2P$ 
electron, and that with the even parity is explained as the 
compound state of two $2P$ electrons. The observed energy 
is $-E_{10}^{obs}=2.1332(e^{2}/a_{0}$)=58.05eV. The relative 
error for the calculated energy $-E^{para}_{10}$ 
is 4 \%. 

\section{DISCUSSIONS}

In this paper we have presented an improved hyperspherical 
harmonic method for directly solving the quantum three-body 
problem. Applying this method to the problem of the helium 
atom, we have calculated the lowest energies 
and the wavefunctions of some states of the $S$ wave and the 
$P$ wave in the parahelium and in the orthohelium. The relative 
error for the energy of the ground state of the parahelium is
in the thousandths, but the remaining results have the
relative error of a few percents. 

The main approximation, which we have made in our calculation,
is to take the finite $N_{j}$ in the series for the wavefunctions. 
If the coefficients of the truncated terms in the series are much smaller
than the needed precision, the calculated results will be satisfactory.
For example, for the ground state of the helium, when $N_{1}=7$, 
$N_{2}=8$ and $N_{3}=16$, we find that all the coefficients at 
the truncated terms are much less than 0.001:
$$f_{7,0,0}=1.046\times 10^{-7},~~~~~ 
f_{0,8,0}=2.995\times 10^{-4},~~~~~ 
f_{0,0,16}=2.983\times 10^{-4}.\eqno (65) $$

\noindent
Therefore, the relative error of the calculated energy must
be in the thousandths. It also can be seen in the calculated 
results of the energy when different $N_{j}$ are taken:

$$-E_{00}=\left\{\begin{array}{ll} 2.88331(e^{2}/a_{0}),~~~~~{\rm when}~~
N_{1}=5,~~N_{2}=4,~~N_{3}=8, \\ 
2.88805(e^{2}/a_{0}),~~~~~{\rm when}~~
N_{1}=5,~~N_{2}=6,~~N_{3}=12,\\
2.88935(e^{2}/a_{0}),~~~~~{\rm when}~~
N_{1}=5,~~N_{2}=8,~~N_{3}=16,\\
2.88935(e^{2}/a_{0}),~~~~~{\rm when}~~
N_{1}=7,~~N_{2}=8,~~N_{3}=16,\\
2.88972(e^{2}/a_{0}),~~~~~{\rm when}~~
N_{1}=7,~~N_{2}=8,~~N_{3}=25.
\end{array}    \right. \eqno (66) $$

\noindent
Note that, when $N_{2}=8$ and $N_{3}=16$, the calculated 
energy when $N_{1}=5$ is the same as that when $N_{1}=7$.
From those data we may conclude that the finite $N_{j}$ we 
have taken for the calculation of the energy of the ground 
state of the helium atom are suitable for the relative 
precision in the thousandths. 

From the viewpoint of calculation, to raise $N_{1}$ and $N_{3}$
will only increases the calculating quantity, but when $N_{2}$ 
raises, the formula for calculating the coefficients 
$D(n,m,n',m')$ becomes much complicated. Now, it is hard for 
us to raise $N_{2}$ more than eight. It is the reason that we 
have to accept the precision of a few percents for the remaining 
calculated energies, and to leave the improvement of the precision
in the future. For example, 
for the lowest-energy state of the $S$ wave in the orthohelium, 
when $N_{1}=7$, $N_{2}=8$ and $N_{3}=17$, the coefficients 
at the truncated terms are
$$g_{7,0,1}=3.430\times 10^{-4},~~~~~ 
g_{0,8,1}=1.663\times 10^{-3},~~~~~ 
g_{0,0,17}=1.650\times 10^{-3}.\eqno (67) $$

\noindent
and the calculated energies are as follows when different 
$N_{j}$ are taken:
$$-E_{00}^{ortho}=\left\{\begin{array}{ll} 
2.06704(e^{2}/a_{0}),~~~~~{\rm when}~~
N_{1}=5,~~N_{2}=6,~~N_{3}=13,\\
2.06762(e^{2}/a_{0}),~~~~~{\rm when}~~
N_{1}=7,~~N_{2}=6,~~N_{3}=13,\\
2.07932(e^{2}/a_{0}),~~~~~
{\rm when}~~N_{1}=5,~~N_{2}=8,~~N_{3}=17, \\ 
2.08039(e^{2}/a_{0}),~~~~~{\rm when}~~
N_{1}=7,~~N_{2}=8,~~N_{3}=17, \\
2.08396(e^{2}/a_{0}),~~~~~{\rm when}~~
N_{1}=7,~~N_{2}=8,~~N_{3}=25.
\end{array}    \right. \eqno (68) $$

\noindent
The relative error is a few percents.

We would like to pay more attention to the lowest-energy 
state of the $P$ wave with the even parity in the parahelium. 
In order to show the property of the coefficient $h_{p,n,m}$ in 
the series (52) as subscripts increase, we list 
some coefficients $h_{p,n,m}$ when $N_{1}=7$, $N_{2}=8$, and 
$N_{3}=17$:
$$h_{0,0,1}=1, ~~~h_{1,0,1}=-0.4015, ~~~h_{2,0,1}=0.1362, 
~~~h_{3,0,1}=-0.0446,$$
$$h_{4,0,1}=0.0153, ~~~h_{5,0,1}=-0.0055, ~~~h_{6,0,1}=0.0019, 
~~~h_{7,0,1}=-0.0006,$$
$$h_{0,1,1}=-0.3583, ~~~h_{0,2,1}=0.1460, 
~~~h_{0,3,1}=-0.0692, ~~~h_{0,4,1}=0.0374,$$
$$h_{0,5,1}=-0.0227, ~~~h_{0,6,1}=0.0152, 
~~~h_{0,7,1}=-0.0117, ~~~h_{0,8,1}=0.0086,$$
$$h_{0,0,2}=0.0501, ~~~h_{0,0,3}=-0.3675, 
~~~h_{0,0,4}=-0.0104, ~~~h_{0,0,5}=0.1460,$$
$$h_{0,0,6}=0.0038, ~~~h_{0,0,7}=-0.0679 
~~~h_{0,0,8}=-0.0013, ~~~h_{0,0,9}=0.0352,$$
$$h_{0,0,10}=0.0006,~~~h_{0,0,11}=-0.0201,
~~~h_{0,0,12}=-0.0003,~~~h_{0,0,13}=0.0122,$$
$$h_{0,0,14}=0.0002,~~~h_{0,0,15}=-0.0078,
~~~h_{0,0,16}=-0.0001,~~~h_{0,0,17}=0.0051, \eqno (69) $$

\noindent
We also list some calculated results for the energy when $N_{2}=8$:
$$-E^{para}_{11}=\left\{\begin{array}{ll}
0.576911\left(e^{2}/a_{0}\right) &{\rm when~~}N_{1}=5,{\rm ~~and~~}N_{3}=9\\
0.579576\left(e^{2}/a_{0}\right) &{\rm when~~}N_{1}=5,{\rm ~~and~~}N_{3}=13\\
0.580345\left(e^{2}/a_{0}\right) &{\rm when~~}N_{1}=5,{\rm ~~and~~}N_{3}=17\\
0.577204\left(e^{2}/a_{0}\right) &{\rm when~~}N_{1}=7,{\rm ~~and~~}N_{3}=9\\
0.580270\left(e^{2}/a_{0}\right) &{\rm when~~}N_{1}=7,{\rm ~~and~~}N_{3}=13\\
0.581291\left(e^{2}/a_{0}\right) &{\rm when~~}N_{1}=7,{\rm ~~and~~}N_{3}=17\\
0.581840\left(e^{2}/a_{0}\right) &{\rm when~~}N_{1}=7,{\rm ~~and~~}N_{3}=25. 
\end{array} \right. \eqno (70) $$

\noindent
Due to the restriction in $N_{2}$, the relative error of the 
calculated energy is in the hundredths. However, the existence 
of a $P$ wave state of even parity in the parahelium with the 
energy $-E\sim 0.58 (e^{2}/a_{0})$ is reliable. Note that
this state is forbidden for the electric dipole transition 
to the ground state, but it is allowed to the $P$ wave state 
with odd parity. The energy difference in the allowed transition 
is around 39.18eV. We are waiting for the observation of this 
transition in experiments. 

It seems that the existence of this state conflicts with the shell 
model in the atomic physics. In the shell model two electrons in 
the helium atom are supposed to fill in the energy levels of a 
hydrogen-like atom according to the exclusion principle. The
lowest-energy state of the $P$ wave with the even parity has
to be composed of two $P$ wave electrons. From the shell model, 
two $2P$ electrons cannot compose a $P$ wave state with symmetric 
spatial wavefunction, and this state has to be explained as the
compound state of one $2P$ and one $3P$ electrons. In this state
each electron moves in the electric field of the nucleus, 
screened by the other electron:
$$-E=\displaystyle {Z_{1}^{2}e^{2} \over 2\cdot 2^{2}\cdot a_{0}}
+\displaystyle {Z_{2}^{2}e^{2} \over 2\cdot 3^{2}\cdot a_{0}}. $$

\noindent
If one electron moves completely inside of the other electron,
$Z_{1}=2$ and $Z_{2}=1$, or vice versa. In the real case, 
$Z_{1}$ is less than $2$ and $Z_{2}$ is larger than $1$. We
assume that $Z_{1}=2-\tau$, $Z_{2}=1+\tau$, and $0\leq \tau \leq 1$.
Thus, the energy $-E_{10}=-E(\tau)$ is a function of the parameter $\tau$, 
and given in the unit $e^{2}/a_{0}$ as follows: 
$$-E(0)=0.556,~~~-E(0.1)=0.518,~~~-E(0.2)=0.485,~~~-E(0.3)=0.455,$$
$$-E(0.4)=0.429,~~~-E(0.5)=0.406,~~~-E(0.6)=0.387,~~~-E(0.7)=0.372,$$
$$-E(0.8)=0.360,~~~-E(0.9)=0.352,~~~-E(1)=0.347.~~~~~~~~~~~~~~~~~ $$

\noindent
This estimation is very rough, but it shows qualitatively
that this $P$ wave state seems not to be the compound state
of one $2P$ and one $3P$ electrons in the shell model. If it
is true, the shell model in the atomic physics is challenged 
even qualitatively. We sincerely hope that this $P$ wave state
with the even parity in the parahelium will be studied
further in both theoretical and experimental physics.

\vspace{5mm}
\noindent
{\bf ACKNOWLEDGMENTS}. One of the authors (ZQM) would like to 
thank Prof. Hua-Tung Nieh and Prof. Wu-Yi Hsiang for drawing 
his attention to this problem. The authors are thankful to Profs.
Bo-Yuan Hou and Yi-Fan Liu for helpful discussions. This work 
was supported by the National Natural Science Foundation of 
China and Grant No. LWTZ-1298 of the Chinese Academy of Sciences.

\begin{center}

{\bf Appendix} $~~~$PROOF OF SOME IDENTIFIES
\end{center}

In this appendix we will prove some identities for the
alternating series used in this paper.
$$\displaystyle \sum_{r=0}^{n}~\displaystyle {(-1)^{r}(n+m+r)!
\over r!(n-r)!(m+r)!} =(-1)^{n}. \eqno (A1) $$

\noindent
{\bf Proof}: From Newton's binomial expansion
$$(1+x)^{n}=\displaystyle \sum_{m=0}^{n}~\displaystyle 
{n! x^{m} \over m!(n-m)!}, $$
$$(1+x)^{-n}=\displaystyle \sum_{m=0}^{\infty}~\displaystyle 
{(-1)^{m}(n+m-1)!x^{m} \over m!(n-1)!}, $$

\noindent
we have
$$\displaystyle \sum_{t=0}^{\infty}~(-1)^{t} x^{t} 
=(1+x)^{-1}=(1+x)^{n-n-1}$$
$$~~~~~=\displaystyle \sum_{r=0}^{n}~\displaystyle 
{n! x^{r} \over r!(n-r)!}~\displaystyle \sum_{s=0}^{\infty}~\displaystyle 
{(-1)^{s}(n+s)!x^{s} \over s!n!} $$
$$~~~~~=\displaystyle \sum_{t=0}^{\infty}~(-1)^{t} x^{t} 
~\displaystyle \sum_{r=0}^{n}~\displaystyle {(-1)^{r}(n+t-r)!
\over r!(n-r)!(t-r)!} ,  $$

\noindent
where the summation index $s$ is replaced by $t=r+s$. Comparing two 
sides of the equality, we obtain
$$\displaystyle \sum_{r=0}^{n}~\displaystyle {(-1)^{r}(n+t-r)!
\over r!(n-r)!(t-r)!}=1. $$

\noindent
Letting $t=n+m$, and then replacing the summation index $r$
by $n-r$, we proved Eq. (A1). Q.E.D.

$$F(n,m,t)\equiv \displaystyle \sum_{r=0}^{n}~\displaystyle {(-1)^{r}(n+m+r)!
\over r!(n-r)!(m+r)!(m+t+r)} =
\displaystyle {(-1)^{n} (t-n)_{n}
\over (m+t)_{n+1}}. \eqno (A2) $$

\noindent
where $(a)_{n}=a(a+1)\cdots (a+n-1)$ and $(a)_{0}=1$.

{\bf Proof}: Prove Eq. (A2) by induction. It is obvious that 
Eq. (A2) holds when $n=0$ and $n=1$. Now, assuming that Eq. (A2) 
holds for $n\leq s$, we will prove it holds when $n=s+1$. In fact,
$$F(s+1,m,t)(m+t+s+1)=\displaystyle \sum_{r=0}^{s+1}~\displaystyle 
{(-1)^{r}(s+1+m+r)! \over r!(s+1-r)!(m+r)!} 
\displaystyle {(m+t+r)+(s+1-r)\over (m+t+r)}$$
$$=(-1)^{s+1}+\displaystyle \sum_{r=0}^{s}~\displaystyle 
{(-1)^{r}(s+m+r)! \over r!(s-r)!(m+r)!} 
\displaystyle {(m+t+r)+(s+1-t)\over (m+t+r)}~~~~~~~~~~~~~~~~~~~~~~~~$$
$$=-F(s,m,t)(t-s-1).~~~~~~~~~~~~~~~~~~~~~~~~~~~~~~~~~~~~~~~~~~~~~~~~~~~~  $$

\noindent
Removing the factor $(m+t+s+1)$ from the left to the right,
we proved that Eq. (A2) holds for $n=s+1$. Q.E.D.

$$G(n,m,t,b,\ell)=\displaystyle \sum_{r=0}^{n}~\displaystyle {(-1)^{r}(n+m+r)!
\over r!(n-r)!(m+r)!}\displaystyle  {(m+b+r)_{\ell} \over (m+t+r)_{\ell+1}}$$
$$=\displaystyle {(-1)^{n} (t-n+\ell)_{n-\ell}
\over (m+t)_{n+\ell+1}} \displaystyle \sum_{s=0}^{\ell}~(-1)^{s}
(b-t)_{s}~P^{(\ell)}_{s},  \eqno (A3) $$

\noindent
where
$$P^{(0)}_{0}=1,$$
$$P^{(1)}_{0}=t(m+t),$$
$$P^{(1)}_{1}=-t+n(n+m+1), $$
$$P^{(2)}_{0}=(t+1)t(m+t)(m+t+1),$$
$$P^{(2)}_{1}=2t(m+t)[-(t+1)+n(n+m+1)],$$
$$P^{(2)}_{2}=t[t+1-2n(n+m+1)]+(n-1)n(n+m+2)(n+m+1)/2,$$
$$P^{(3)}_{0}=(t+2)(t+1)t(m+t)(m+t+1)(m+t+2),$$
$$P^{(3)}_{1}=3(t+1)t(m+t)(m+t+1)[-(t+2)+n(n+m+1)],$$
$$P^{(3)}_{2}=3t(m+t)\{(t+1)[t+2-2n(n+m+1)]$$
$$~~~+n(n-1)(n+m+1)(n+m+2)/2\}, $$
$$P^{(3)}_{3}=t(t+1)[-(t+2)+3n(n+m+1)]-3tn(n-1)(n+m+1)(n+m+2)/2$$
$$~~~+n(n-1)(n-2)(n+m+1)(n+m+2)(n+m+3)/6, $$
$$P^{(4)}_{0}=(t+3)(t+2)(t+1)t(m+t)(m+t+1)(m+t+2)(m+t+3),$$
$$P^{(4)}_{1}=4(t+2)(t+1)t(m+t)(m+t+1)(m+t+2)[-(t+3)+n(n+m+1)],$$
$$P^{(4)}_{2}=6(t+1)t(m+t)(m+t+1)\{(t+2)[t+3-2n(n+m+1)]~~~~~~~~~~$$
$$~~~+n(n-1)(n+m+1)(n+m+2)/2\},$$
$$P^{(4)}_{3}=4t(m+t)\{(t+1)(t+2)[-(t+3)+3n(n+m+1)]~~~~~~~~~~~~~~$$
$$~~~-3(t+1)n(n-1)(n+m+1)(n+m+2)/2~~~~~~~~~~$$
$$~~~+n(n-1)(n-2)(n+m+1)(n+m+2)(n+m+3)/6\}, $$
$$P^{(4)}_{4}=t(t+1)(t+2)[t+3-4n(n+m+1)]~~~~~~~~~~~~~~~~~~~~~~~~~$$
$$~~~+3t(t+1)n(n-1)(n+m+1)(n+m+2)~~~~~~~~~~~~~~~~~~~~~~$$
$$~~~-2tn(n-1)(n-2)(n+m+1)(n+m+2)(n+m+3)/3~~~~~~~~~~~~ $$
$$~~~+n(n-1)(n-2)(n-3)(n+m+1)(n+m+2)(n+m+3)(n+m+4)/24, $$

\noindent
The remaining coefficients can be obtained from us upon request.

{\bf Proof}: When $\ell=0$, $G(n,m,t,b,0)=F(n,m,t)$.
When $\ell=1$, we have
$$G(n,m,t,t,1)=\displaystyle {(-1)^{n} (t-n+1)_{n-1}
\over (m+t)_{n+2}}P^{(1)}_{0}=F(n,m,t+1), $$
$$G(n,m,t,t+1,1)=\displaystyle {(-1)^{n} (t-n+1)_{n-1}
\over (m+t)_{n+2}}\left\{P^{(1)}_{0}-P^{(1)}_{1}\right\}=F(n,m,t).  $$

\noindent
$P^{(1)}_{1}$ and $P^{(1)}_{0}$ can be solved from those 
relations. When $\ell=2$, we have
$$G(n,m,t,t,2)=\displaystyle {(-1)^{n} (t-n+2)_{n-2}
\over (m+t)_{n+3}} P^{(2)}_{0}=F(n,m,t+2), $$
$$G(n,m,t,t-1,2)=\displaystyle {(-1)^{n} (t-n+2)_{n-2}
\over (m+t)_{n+3}}\left\{P^{(2)}_{0}+P^{(2)}_{1}\right\}=G(n,m,t+1,t-1,1), $$
$$G(n,m,t,t+1,2)=\displaystyle {(-1)^{n} (t-n+2)_{n-2}
\over (m+t)_{n+3}}\left\{P^{(2)}_{0}-P^{(2)}_{1}+2P^{(2)}_{2}\right\}
=F(n,m,t), $$
$$G(n,m,t,t+2,2)=\displaystyle {(-1)^{n} (t-n+2)_{n-2}
\over (m+t)_{n+3}}\left\{P^{(2)}_{0}-2P^{(2)}_{1}+6P^{(2)}_{2}\right\} $$
$$~~~=G(n,m,t,t+3,1).$$

\noindent
The first three relations are used for calculating the coefficients
$P^{(2)}_{2}$, $P^{(2)}_{1}$ and $P^{(2)}_{0}$, and the last one
is for check. Similarly, the coefficients $P^{(\ell)}_{s}$ can
be calculated from $G(n,m,t,b,\ell')$, where $\ell'<\ell$.

\end{document}